\documentclass[aps,pra,twocolumn,amsmath,amssymb,superscriptaddress,tightenlines]{revtex4}
\usepackage{graphicx}
\usepackage{epsfig}
\usepackage{dcolumn}
\usepackage{bm}
\usepackage{amssymb}
\usepackage{bm}
\usepackage{amsmath, bm}
\usepackage{upgreek}
\usepackage[colorlinks,citecolor=blue,linkcolor=blue,hyperindex]{hyperref}

\begin{document}

\title{Eigenstate thermalization and disappearance of quantum many-body scar states in weakly interacting fermion systems}
\author{Ken K. W. Ma}
\affiliation{National High Magnetic Field Laboratory, Tallahassee, Florida 32310, USA}
\author{A. Volya}
\affiliation{Department of Physics, Florida State University, Tallahassee, Florida 32306, USA}
\author{Kun Yang}
\affiliation{National High Magnetic Field Laboratory, Tallahassee, Florida 32310, USA}
\affiliation{Department of Physics, Florida State University, Tallahassee, Florida 32306, USA}
\date{\today}


\begin{abstract}
The recent discovery of quantum many-body scar states has revealed the possibility of having states with low entanglement that violate the eigenstate thermalization hypothesis in nonintegrable systems. Eigenstates with low entanglement entropy are rare but naturally exist in the integrable system of free fermions. Here, we demonstrate analytically that these atypical states would be always eliminated when an arbitrary weak interaction is introduced between the fermions. In particular, we show that the probability of having a many-body scar state with entanglement entropy satisfying a sub-volume scaling law decreases double exponentially as the system size. Thus, our results provide a quantitative argument for the disappearance of scar states in interacting fermion systems.
\end{abstract}

\maketitle

\section{Introduction}

An important issue in statistical mechanics is its emergence in isolated quantum systems. The eigenstate thermalization hypothesis (ETH) conjectures that highly excited states are thermal (in a sense to be discussed in more detail below) in nonintegrable systems~\cite{Deutsch1991, Srednicki1994, Rigol2008, Srednicki1999, Rigol2016, Deutsch-review}. However, progress (especially analytic) toward establishing ETH is very slow and difficult.

The recent discovery of many-body scar states has revealed the possibility of violation of ETH. More specifically, some systems do not achieve thermalization after a long period of time when they were prepared in some special initial states. Various specific models have been found to have many-body scar states. A common diagnostic of these scar states is their low entanglement entropy, which violates the usual volume law~\cite{Vidmar2021} (for reviews, see~\cite{scar-review1, scar-review2, scar-review3, scar-review4} and the references therein). We will only focus on this kind of scar states which possess sub-extensive entanglement entropy here, although recent work has proposed the concept of ``rainbow scars" that may have extensive EE~\cite{rainbow1, rainbow2}. On the other hand, many-body scar states are forbidden in the strong version of ETH~\cite{strong-ETH, Huse, Grover}. Whether scar states exist or not in generic systems remains a central problem in quantum statistical mechanics, that is of foundational importance.

It is useful to compare the existence of scar states with what happens in free fermion systems, which, while being (trivially) integrable, provide surprising insights to the generic situations, and lay the foundation for the results we report in this paper on weakly interacting fermion systems. It was demonstrated that the overwhelming majority of highly excited free fermion (Fock) states, which were termed typical states, are thermal in the sense that the reduced density matrix $\rho_A$ of a small subsystem $A$ is nothing but the thermal density matrix that gives rise to the same energy and particle densities of the parent state, a property termed {\em eigenstate typicality}~\cite{Yang2015}. An immediate consequence is that their entanglement entropy is identical to the thermal entropy~\cite{footnote-entropy}. The same reference also pointed out that this feature is violated in the so-called atypical eigenstates, which have low entanglement entropy. Our recent work further clarified that the ratio of atypical to typical eigenstates is exponentially small in system size~\cite{Yang2022}. Such atypical states can be viewed as the noninteracting version of many-body scar states. By introducing interaction between fermions, the integrability of the system is broken. A simple perturbative argument for weakly interacting systems suggests that a generic eigenstate is a mix between the free fermion eigenstates with similar kinetic energies. Due to the overwhelming majority of typical free fermion eigenstates, it is conjectured that atypical ones will be washed out in the mixing process~\cite{Yang2018}, hence it is highly unlikely to have scar states with low entanglement in the interacting fermion system. As we will demonstrate below, this is indeed the case, and the full knowledge of the free fermion eigenstates is crucial for studying thermalization and many-body scar states in the interacting system. 

\section{Thermalization in fermion systems with two-body interaction} 

We now review existing results on thermalization in fermion systems with two-body interaction. Different from previous literature which were mostly in the context of nuclear physics, our summary below uses the convention in condensed matter physics instead, in which the thermodynamic limit is important.

We define the average energy spacing between single particle states as $\Delta$. It is common to set $\Delta=1$ in nuclear physics as the size of a nucleus and the number of nucleons are fixed. However, we do not follow this convention here since our discussion will focus on the thermodynamic limit. This limit is defined as having both number of fermions $n$, and the number of single particle states $N$ (proportional to the system size) going to infinity, with the ratio $\nu=n/N$ fixed. As a result, the scaling behaviors of physical quantities in $N$ or $n$ are important. With this in mind, we keep $\Delta$ explicit in the following discussion. This quantity is inversely proportional to the volume of the system, i.e., $\Delta\sim W/N$ where $W$ is the (single particle) band width. It provides a unit to measure different energy scales in the system. Note that we use $\sim$ to denote scaling behavior.

The system of fermions with two-body interaction is described by the Hamiltonian,
\begin{eqnarray} \label{eq:Hamiltonian}
H
=H_0
+\lambda\sum_{\substack{i<j, r<s \\ (r,s)\neq (i,j)}}
\langle r,s |V_2| i,j \rangle
f_s^\dagger f_r^\dagger f_if_j,
\end{eqnarray}
where $H_0=\sum_i \epsilon_i f_i^\dagger f_i$ is the free fermion Hamiltonian. All subscripts $i,j,r,s$ label the single particle states. Each of them contains both orbital and spin indices, which are collectively denoted by an index that can take value from $1$ to $N$. The second quantized fermionic operators satisfy $\{f_i, f_j^\dagger\}=\delta_{ij}$. Furthermore, we assume the system preserves translational symmetry, such that the single particle states are plane-wave states labeled by momenta and spins. The many-body eigenstates of $H_0$, denoted collectively as $|\psi_a\rangle$, are given by the $M=N!/n!(N-n)!=2^{Nh(\nu)}$ different Slater determinants of the single particle states.  Here,
\begin{eqnarray}
h(\nu)=-\nu\log_2 \nu -(1-\nu)\log_2 (1-\nu).
\end{eqnarray}  
The residual two-body interaction (off-diagonal matrix elements) is denoted as $\lambda V_2$. Note that the interaction strength is solely characterized by the dimensionful parameter $\lambda$, which scales as $U/N$, where $U$ is a Hubbard-like (local) interaction strength. For the Hubbard model, the total spin of the system is conserved. The matrix elements $\langle r,s |V_2| i,j \rangle$ take $O(1)$ values that are independent of the system size. As long as the interaction is short-ranged, the details are irrelevant to our discussions below. Using $|\psi_a\rangle$, a generic eigenstate of $H$ can be written as
\begin{eqnarray}
|\Psi_\alpha\rangle
=\sum_{a=1}^M
c^\alpha_a |\psi_a\rangle.
\end{eqnarray}
For real symmetric $V_2$, all $c^\alpha_a$ are real numbers.

In the leading-order contribution, $V_2$ only couples $|\psi_a\rangle$ and $|\psi_b\rangle$ that differ by at most two single particle states. For a given $|\psi_a\rangle$, the number of nonzero matrix elements for $V_2$ is
\begin{eqnarray}
K_2
=n(N-n)+\frac{n(n-1)(N-n)(N-n-1)}{4}.
\end{eqnarray}
In the thermodynamic limit, $K_2\approx n^2 N^2/4\ll M$ when $\nu$ is not very close to zero or one. Therefore, $V_2$ is a very sparse matrix when it is embedded in the entire Hilbert space~\cite{French, French2, Bohigas, Kota2001, Kota-book}. This feature suggests that the results from the usual random matrix theory should not be naively generalized for describing systems with two-body interaction. 

For finite systems, increasing $\lambda$ from zero would lead to various crossovers in the distribution of eigenenergies and the structure of eigenfunctions for the interacting fermion system~\cite{Flambaum, Flambaum2, Flambuam3, Flambuam4, Kota2001, Kota-book, Zelevinsky-rev, Aberg1, Aberg2, Shepelyansky1, Shepelyansky2, Zelevinsky, Kota2004, Kota2011, Gopalakrishnan2022}. This can be understood by comparing different energy scales in the system.

Two states $|\psi_a\rangle$ and $|\psi_b\rangle$ that $V_2$ can directly couple have a maximum energy difference $\Delta_2\sim N\Delta$. The typical density of directly coupled states via $V_2$ is estimated as 
$\rho_2=K_2/\Delta_2\sim n^2 N/\Delta$. The energy levels start to mix and avoid crossing each other when $\lambda$ becomes comparable to $\rho_2^{-1}$. Hence, the energy level spacing obeys the Wigner-Dyson distribution when $\lambda\geq\lambda_c$~\cite{Shepelyansky1}, where
\begin{eqnarray} \label{eq:lambda_c}
\lambda_c
\sim\rho_2^{-1}
\quad\Rightarrow\quad
\frac{\lambda_c}{\Delta}\sim\frac{1}{n^2 N}.
\end{eqnarray}

Instead of the Wigner-Dyson distribution in energy level spacing, the structure of eigenfunctions plays a more important role in thermalization. More specifically, achieving thermalization (or satisfying the ETH) requires the eigenstates of the system to be sufficiently chaotic or delocalized in Fock space. This is realized when the eigenstate can be viewed as a random superposition of $|\psi_a\rangle$ in an energy shell depending on $V_2$~\cite{Izrailev2012-PRL, Izrailev2012-PRE, Izrailev2017-AIP, Izrailev2017-PRE, Zelevinsky}. The delocalization of $|\Psi_\alpha\rangle$ in the space of basis states $|\psi_a\rangle$ can be studied from the following spreading function,
\begin{eqnarray}
F^{\alpha}(E)
=\sum_{a=1}^M |c^\alpha_a|^2~\delta(E-E_a).
\end{eqnarray}
The shape of $F^\alpha(E)$ is very similar to the shape of the strength function, which is defined as~\cite{Sasha-book}
\begin{eqnarray}
F_a(E)
=\sum_{\alpha=1}^M |c^\alpha_a|^2~\delta(E-E^\alpha).
\end{eqnarray}

Similar to the distribution of energy level spacing, both $F^\alpha(E)$ and $F_a(E)$ undergo crossovers when $\lambda$ is increased from zero. Since $M$ is exponentially large in the system size, the average energy spacing for the eigenstates of $H_0$ is exponentially small. Hence, the strength function takes a Lorentzian (or Breit-Wigner) distribution for an exponentially small $\lambda>\lambda_0$~\cite{Kota-book}. By further increasing $\lambda$, the strength function changes from the Lorentizian to the Gaussian distribution~\cite{Flambaum, Flambuam4, Kota2001, Aberg1, Varga2002, Izrailev2006}. Previous work~\cite{Varga2002, Kota-book} suggested that this crossover occurs when $\lambda$ exceeds
\begin{eqnarray} \label{eq:lambda-F}
\frac{\lambda_F}{\Delta}\sim \frac{1}{\sqrt{N}}.
\end{eqnarray}
The critical values satisfy $\lambda_0\ll \lambda_c<\lambda_F$~\cite{footnote-Fock}.

Beyond the Gaussian regime (i.e., $\lambda>\lambda_F$), the strength function takes the form,
\begin{eqnarray}
F_a(E)
=\frac{1}{\sqrt{2\pi}\sigma_a}
\exp{\left[-\frac{(E-\bar{E}_a)^2}{2\sigma_a^2}\right]}.
\end{eqnarray}
The variance $\sigma_a^2$ is given by
\begin{eqnarray}
\sigma_a^2
=\langle\psi_a|H^2|\psi_a\rangle-\left(\langle\psi_a|H|\psi_a\rangle\right)^2.
\end{eqnarray}
In the middle of the many-body energy spectrum, $\sigma_a^2$ is independent of $a$~\cite{Kota-book}. It is entirely determined by the off-diagonal matrix elements of $H$, namely the residual two-body interaction $\lambda V_2$. Different results for $\sigma_a^2$ were reported in previous literature~\cite{Flambuam4, Kota-book}, but their asymptotic forms agree in the thermodynamic limit,
\begin{eqnarray}
\sigma_a^2
\approx \frac{\lambda^2}{4}n^2 N^2\left(1-\frac{n}{N}\right)^2.
\end{eqnarray}

Although $|\Psi_\alpha\rangle$ contains a large number of basis states $|\psi_a\rangle$ when $\lambda>\lambda_F$, the eigenstates actually do not delocalize in the entire $M$-dimensional Hilbert space~\cite{Izrailev2012-PRL, Izrailev2012-PRE, Izrailev2017-AIP, Izrailev2017-PRE}. Instead, the basis states contained in $|\Psi_\alpha\rangle$ span an energy shell that is determined by the effective bandwidth $\sigma_a$. Within this energy shell, the number of many-body basis states is roughly estimated as
\begin{eqnarray} \label{eq:eff-xi}
M'
\approx\frac{M}{\Delta_n}\sigma_a
\sim \frac{M\lambda}{\Delta}
\sim M\left(\frac{U}{W}\right).
\end{eqnarray}
Here, $\Delta_n\sim n(N-n)\Delta$ is the bandwidth of the many-body energy spectrum. From Eq.~\eqref{eq:eff-xi}, one has $1\ll M'\ll M$ for a weak interaction $\lambda\ll\Delta$. In the thermodynamic limit, $|\Psi_\alpha\rangle$ only delocalizes in a restricted Hilbert space which is a fraction of the entire Hilbert space. Now, $|\Psi_\alpha\rangle$ has real random coefficients $c^\alpha_a$ along the basis states $|\psi_a\rangle$ restricted in the energy shell~\cite{Zelevinsky}. 

In the thermodynamic limit, the above discussion indicates that an arbitrary small but nonzero interaction between the fermions always leads to an energy shell, in which $|\Psi_\alpha\rangle$ is a random superposition of $|\psi_a\rangle$. From this, one may argue that thermalization is achieved in the thermodynamic limit, based on the notion of canonical typicality~\cite{Goldstein-CT, Popescu2006}. The canonical typicality states that the reduced density matrix $\rho_A$ of a small subsystem $A$ generated from $|\Psi_\alpha\rangle$ approaches that generated from the microcanonical ensemble with probability one (more details in later discussion). This is quite similar to eigenstate typicality~\cite{Yang2015}, and corresponds to the local version of ETH termed subsystem ETH~\cite{Liu-ETH18, CFT-sETH}. However, canonical typicality does not eliminate the possibility of having scar states in the system. As we mentioned before, the strong version of ETH is a much stronger statement that forbids the existence of any many-body scar state in the nonintegrable system, and their existence impedes thermalization of certain specially prepared initial states. Thus to fully understand thermalization of weakly interacting fermion systems, especially whether the (strong version of) ETH holds or not, require us to go beyond the existing results and study if scar states (which can be viewed as the interacting or nonintegrable version of atypical free fermion states) survive weak interaction. 

\section{Elimination of many-body scar states via typicalities} 

In this section, we study the following question. What is the probability for a random eigenstate $|\Psi_\alpha\rangle$ to have its entanglement entropy violating the usual volume law? It is important to emphasize that the entanglement entropy of the eigenstate $|\Psi_\alpha\rangle$ is not necessarily larger than the entanglement entropy of any one of the basis states $|\psi_a\rangle$~\cite{Smolin-EE, Gour2008}. Therefore, eigenstate typicality in free fermion system does not immediately eliminate the possibility of having many-body scar states in the interacting system. Instead, it is necessary to study the reduced density matrix,
\begin{eqnarray} \label{eq:RDM-Psi}
\text{Tr}_B \rho_\Psi=\text{Tr}_B(|\Psi_\alpha\rangle \langle\Psi_\alpha|),
\end{eqnarray}
and its corresponding von Neumann entropy. We define states whose entanglement entropy has sub volume-law scaling as a scar state~\cite{scar-review1, scar-review2, scar-review3, scar-review4}. As a result, the aforementioned probability tells us the expected number of scar states in a particular system.

The whole system without any restriction from symmetry can be decomposed into two subsystems $A$ and $B$, which have their respective Hilbert spaces $\mathcal{H}_A$ and $\mathcal{H}_B$. Here, $\mathcal{H}_F=\mathcal{H}_A\otimes\mathcal{H}_B$, where $\mathcal{H}_F$ stands for the entire Fock space of fermions that has the dimension $M_F=2^N$. In addition, $M_A=2^{N_A}$ and $M_B=2^{N_B}$ are the dimensions of $\mathcal{H}_A$ and $\mathcal{H}_B$, respectively. We fix the size of subsystem $A$, whereas the size of subsystem $B$ goes to infinity in the thermodynamic limit. Thus, $1\ll M_A\ll M_B$. We further restrict the allowed states of the system to the subspace $\mathcal{H}_R\subset\mathcal{H}_F$, where $\mathcal{H}_R$ is spanned by the $M'$ free fermion eigenstates in the energy shell. Hence, its dimension satisfies $1\ll M'\ll M<M_F$. It is emphasized that $|\Psi_\alpha\rangle$ can be defined as a random pure state in $\mathcal{H}_R$, but not in $\mathcal{H}_F$ or $\mathcal{H}$ which has the dimension $M$.

Following the original work on canonical typicality~\cite{Popescu2006, Goldstein-CT}, we start by studying the trace distance between $\text{Tr}_B \rho_\Psi$ and another reduced density matrix $\text{Tr}_B \rho_d$. Here,
\begin{eqnarray} \label{eq:rho-d}
\rho_d
=\frac{1}{M'}\mathbb{I}_{M'\times M'}
=\frac{1}{M'}\sum_{a=1}^{M'}|\psi_a\rangle\langle\psi_a|.
\end{eqnarray}
The notation $\mathbb{I}_{M'\times M'}$ stands for the $M'\times M'$ identity matrix in $\mathcal{H}_R$~\cite{footnote-embedding}. The trace distance between $\text{Tr}_B \rho_\Psi$ and $\text{Tr}_B \rho_d$ is defined as~\cite{Chuang-book, Wilde-book}
\begin{align}
\nonumber
&T(\text{Tr}_B \rho_\Psi, \text{Tr}_B \rho_d)
\\
=~&\frac{1}{2}~\text{Tr}_A
\left[
\sqrt{\left(\text{Tr}_B \rho_\Psi -\text{Tr}_B \rho_d\right)^\dagger
\left(\text{Tr}_B \rho_\Psi -\text{Tr}_B \rho_d\right)}
\right].
\end{align}

Now, consider a function $f(\mathbf{x})$, where $\mathbf{x}$ is a point chosen uniformly at random from the unit hypersphere $S^d$. Here $d$ needs to be large. Then, Levy's lemma states that
\begin{eqnarray}
\mathbb{P}
\left(|f(\mathbf{x})-\langle f\rangle|\geq\epsilon\right)
\leq 2\exp{\left[-\frac{(d+1)\epsilon^2}{9\pi^3\eta^2}\right]}.
\end{eqnarray}
The symbol $\eta$ denotes the Lipschitz constant of the function $f$, which is defined as $\eta=\sup|\nabla f|$. The lemma states that the probability for $f(\mathbf{x})$ to deviate from its expectation value by a large amount is exponentially small.

For the function $f=T(\text{Tr}_B \rho_\Psi, \text{Tr}_B \rho_d)$, $\eta\leq 1$ was found in~\cite{Popescu2006}. In the present case, we have $d=M'-1$. Following the approach in Ref.~\cite{Popescu2006}, one obtains
\begin{align} \label{eq:CT}
\mathbb{P}\left(T(\text{Tr}_B \rho_\Psi, \text{Tr}_B \rho_d)
\geq\epsilon+\frac{1}{2}\sqrt{\frac{M_A}{M_{B,\text{eff}}}}~ \right)
\leq 2e^{-M'\epsilon^2/9\pi^3}.
\end{align}
Note that $\langle f\rangle=(1/2)\sqrt{M_A/M_{B,\text{eff}}}$ in the present case. Also, $M_{B,\text{eff}}=1/\text{Tr}_B [\left(\text{Tr}_A\rho_\Psi\right)^2]\geq M'/M_A$ is the effective dimension of the subsystem $B$. If one takes $M_{B,\text{eff}}=M'/M_A$, then Eq.~\eqref{eq:CT} becomes
\begin{align} \label{eq:CT-GOE}
\mathbb{P}
\left(
T(\text{Tr}_B \rho_\Psi, \text{Tr}_B \rho_d)
\leq \epsilon+\frac{M_A}{2\sqrt{M'}}~ \right)
\geq 1-2e^{-M'\epsilon^2/9\pi^3}.
\end{align}
In the thermodynamic limit we have $M_A/\sqrt{M'}\rightarrow 0$ and can take $\epsilon\ll 1 \ll M'\epsilon^2$, to find the probability for $T(\text{Tr}_B \rho_\Psi, \text{Tr}_B \rho_d)$ below a vanishing upper bound tending to one. Note that $M'\sim M(\lambda/\Delta)\rightarrow\infty$ in the thermodynamic limit. Eq.~\eqref{eq:CT-GOE} is the mathematical statement of canonical typicality~\cite{Goldstein-CT}.

Since $\text{Tr}_B \rho_\Psi$ and $\text{Tr}_B \rho_d$ have a small trace distance, it is expected that the difference between their von Neumann entropy, $\Delta S=|S(\text{Tr}_B \rho_\Psi)-S(\text{Tr}_B \rho_d)|$, is also small. Mathematically, this is guaranteed by the Fannes-Audenaert inequality~\cite{Wilde-book}:
\begin{align} \label{eq:FA-inequality}
\Delta S
\leq T(\text{Tr}_B \rho_\Psi, \text{Tr}_B \rho_d)
\log_2{(M_A-1)}
+C.
\end{align}
We define $C=h\left(T(\text{Tr}_B \rho_\Psi, \text{Tr}_B \rho_d)\right)$, which satisfies $C\leq 1$. From Eqns.~\eqref{eq:CT-GOE} and~\eqref{eq:FA-inequality}, we obtain
\begin{align} \label{eq:EE-bound}
\mathbb{P}\left[\Delta S
<\left(\epsilon+\frac{M_A}{2\sqrt{M'}}\right)\log_2{M_A}+C\right]
\geq
1-2e^{-M'\epsilon^2/9\pi^3}.
\end{align}

For a fixed value of $\lambda/\Delta$, Eq.~\eqref{eq:eff-xi} suggests that $M'$ should also grow exponentially in the system size. Since the size of subsystem $A$ is fixed and being much smaller than the size of the entire system, one has
$M_A\log_2 M_A\ll\sqrt{M'}$. Then, Eq.~\eqref{eq:EE-bound} is reduced to
\begin{align} \label{eq:EE-typicality}
\mathbb{P}\left(\Delta S<\epsilon N_A+C\right)
\geq 1-2\exp{\left(-\frac{M'\epsilon^2}{9\pi^3}\right)}.
\end{align}
When $\epsilon\rightarrow 0$, the entanglement entropy of the system becomes arbitrarily close to $S(\text{Tr}_B\rho_d)$, within a constant term $C\leq 1$.

The above discussion highlights the importance of studying the term $S(\text{Tr}_B\rho_d)$ in greater detail. From Eq.~\eqref{eq:rho-d} and the concavity of von Neumann entropy, $S(\text{Tr}_B\rho_d)$ satisfies the following inequality~\cite{Chuang-book}:
\begin{align} \label{eq:concave}
S\left[\frac{1}{M'}\sum_{a=1}^{M'} \text{Tr}_B\left(|\psi_a\rangle \langle \psi_a|\right)\right]
\geq
\frac{1}{M'}
\sum_{a=1}^{M'}
S\left[\text{Tr}_B\left(|\psi_a\rangle \langle \psi_a|\right)\right].
\end{align}
Notice that $S[\text{Tr}_B(|\psi_a\rangle\langle\psi_a|)]$ is precisely the bipartite entanglement entropy of $|\psi_a\rangle$ in position space. For eigenstates of a free fermion system (i.e., $\lambda=0$ in the present work) with translational symmetry, Refs.~\cite{Yang2015, Yang2022} demonstrated that most of the $|\psi_a\rangle$ have their entanglement entropy identical to the thermal entropy. The volume law $S[\text{Tr}_B(|\psi_a\rangle\langle\psi_a|)]\sim N_A$ is typically satisfied. Therefore,
\begin{eqnarray} \label{eq:Sd-min}
S(\text{Tr}_B \rho_d)
\geq
\frac{1}{M'}\sum_{a=1}^{M'}
S\left[\text{Tr}_B\left(|\psi_a\rangle \langle \psi_a|\right)\right]
=\kappa N_A,
\end{eqnarray}
where $\kappa\leq 1$ is a proportionality constant which does not scale with the system size.

We now denote the entanglement entropy of $|\Psi_\alpha\rangle$ as $\mathcal{E}(\Psi_\alpha)$. From Eqs.~\eqref{eq:EE-typicality}-\eqref{eq:Sd-min}, we obtain
\begin{align} \label{eq:main}
\mathbb{P}
\left[
\mathcal{E}(\Psi_\alpha)
\geq
(\kappa-\epsilon) N_A-C
\right]
\geq 1-2\exp{\left(-\frac{M'\epsilon^2}{9\pi^3}\right)}.
\end{align}
This is the main result in the present work. To violate the volume law in $\mathcal{E}(\Psi_\alpha)$, it requires $\epsilon=\kappa-o(1)$. The upper bound for the corresponding probability vanishes as $\exp{(-c M')}$, with $c=\epsilon^2/9\pi^3\approx \kappa^2/9\pi^3$ being a positive coefficient that is independent of system size. A rough estimate for the expected number of many-body scar states in the energy shell scales as $M'\exp{(-cM')}$. This goes to zero when $M'\rightarrow\infty$ in the thermodynamic limit. The above result suggests that an arbitrary small but nonzero two-body interaction will basically eliminate \textit{all} states with low entanglement. In combination with the arbitrary closeness between $\text{Tr}_B\rho_\Psi$ and $\text{Tr}_B\rho_d$, our result provides a quantitative support for the strong version of ETH in the weakly interacting fermion system.

Although our result looks similar to Theorem III.3 in Ref.~\cite{Hayden2006}, the theorem there was derived for random pure states in a Hilbert space that has a tensor product structure. This structure is absent in $\mathcal{H}_R$. Thus, the theorem does not apply here. Also, $\mathcal{H}_R$ is determined precisely by the constraints in the problem, so the generic result on entanglement in random subspaces is irrelevant. The above comparison clarifies the originality of our work, and further emphasizes the importance of both canonical and eigenstate typicalities in deriving our result.

Furthermore, we should clarify that our discussion has ignored the constraints imposed by the conservation of total momentum and spin in the system. When this issue is taken into account, the eigenstate $|\Psi_\alpha\rangle$ should be a random superposition of free fermion eigenstates with the same total momentum and spin that are also within the energy shell determined by $\lambda V_2$. The number of these basis states still scales exponentially as the system size in the thermodynamic limit. Note that canonical typicality in translationally invariant system was discussed in~\cite{CT-translational}. From both canonical typicality and eigenstate typicality, all (in the probabilistic sense) many-body scar states will be still eliminated in the thermodynamic limit.

One may wonder how these highly unlikely scar states can exist if the above argument looks so general? Suppose the system has an extra symmetry, such as a spectrum generating algebra or Krylov subspaces. Then, there exists a specific set of eigenstates which are not given by random superpositions of $|\psi_a\rangle$ in the restricted energy shell. Consequently, canonical typicality is violated in these rare scar states, so that their entanglement entropy needs not be bounded from below by Eq.~\eqref{eq:main}. This leads to the possibility of having states with low entanglement (or scar states) in the system, hence a violation of the strong version of ETH~\cite{strong-ETH, Huse, Grover, footnote-weakETH, footnote-Katsura}. 

\section{Conclusion and discussion} 

To summarize, we have studied quantitatively the absence of many-body scar states with sub-extensive bipartite entanglement entropy in the weakly interacting fermion system. We show that the probability of having these states vanishes double exponentially in the system size. Therefore, the expected number of such a kind of scar states goes to zero in the thermodynamic limit, albeit the exponentially large number of eigenstates in the system. This occurs whenever there is an arbitrary weak but nonzero interaction between the fermions. On one hand, our work is reminiscent of previous work on integrability breaking~\cite{int1, int2, int3, int4, int5} and the stability of quantum many-body scars under perturbation~\cite{breaking1, breaking2, breaking3, breaking4, breaking5, breaking6}. On the other hand, we have studied directly the common feature of bipartite entanglement entropy of the interacting fermion eigenstates, thus the result in Eq.~\eqref{eq:main} is general, and provides support for the strong version of ETH in the interacting fermion system. 

An important starting point of our work is that a generic eigenstate of the system is a random superposition of Fock states within a small energy shell. We did not take this for granted by assuming the Hamiltonian is a random matrix (as in the usual justification of ETH), which, as discussed earlier, is highly unphysical. Instead, we justified our assumption by extrapolating the existing results in embedded random matrix theory to the thermodynamic limit. By doing so, we can properly identify the energy shell, in which the eigenstates are truly delocalized. This represents an important advancement in justifying ETH.

Last but not least, our work suggests that the absence of scar states relies on both canonical typicality and eigenstate typicality of the free fermion (Fock) basis states, as the entanglement entropy for the interacting eigenstates cannot be deduced from the former alone. This is because the entanglement entropy depends on not only the random nature, but also the properties of the basis states that contribute to the superposition that make up the eigenstates of the interacting system. This highlights the importance of studying entanglement properties of the basis states. However, not many analytical results have been achieved in this direction. The free fermion system provides a good example, in which eigenstate typicality has been clearly demonstrated~\cite{Yang2015}. Although the free fermion system appears very simple and is trivially integrable, it does have non-trivial entanglement, and provides a good starting point for exploring different questions in quantum statistical mechanics in weakly interacting systems. Our results can be generalized to other weakly perturbed intergrable systems, which satisfy both canonical and eigenstate typicalities. This leads us to the conjecture: The strong version of ETH holds in generic interacting systems, unless the system possesses any extra symmetry such as a spectrum generating algebra or Krylov subspaces~\cite{scar-review1, scar-review2, scar-review3, scar-review4}.

\begin{acknowledgments}

This research is supported by the National Science Foundation Grant No. DMR-1932796, 
and by the U.S. Department of Energy Office of Science under Award Number DE-SC0009883. Most of this work was performed at the National High Magnetic Field Laboratory, which is supported by National Science Foundation Cooperative Agreement No. DMR-1644779, and the State of Florida.

\end{acknowledgments}


\begin{thebibliography}{99}
\bibitem{footnote-embedding}
While $\mathbb{I}_{M'\times M'}$ is the identity matrix defined in the subspace $\mathcal{H}_R$, the density matrix $\rho_d$ becomes an $M_F\times M_F$ matrix after embedding $\mathcal{H}_R$ in the original Fock space $\mathcal{H}_F$. In this enlarged Hilbert space, $\rho_d$ takes a block-diagonal form. Similarly, $\rho_\Psi$ is also an $M_F\times M_F$ block-diagonal matrix in $\mathcal{H}_F$. The embedding procedure restores the bipartite nature of the system that enables us to partial tracing out subsystem $B$.
\bibitem{footnote-weakETH}
Note that a small number (with the fraction vanishes in the thermodynamic limit) of scar states is allowed in the weak version of ETH~\cite{strong-ETH, Huse, Grover}.
\end{thebibliography}
\end{document}